\documentclass[aps,showpacs,preprintnumbers,amsmath,amssymb]{revtex4}

\usepackage{graphics,epsfig}
\usepackage{graphicx}
\usepackage{dcolumn}
\usepackage{bm}

\newcommand{\be}{\begin{equation}}
\newcommand{\ee}{\end{equation}}
\newcommand{\ba}{\begin{eqnarray}}
\newcommand{\ea}{\end{eqnarray}}

\begin{document}
\vspace*{2.0cm}
\title{{\bf Asymptotic quasinormal modes of a coupled scalar field in the
 Garfinkle-Horowitz-Strominger dilaton spacetime}}
\author{Songbai Chen}\email{csb3752@hotmail.com}
\author{Jiliang Jing} \email{jljing@hunnu.edu.cn}
\affiliation{ Institute of Physics and  Department of Physics, \\
Hunan Normal University,  Changsha, Hunan 410081, P. R. China }

\vspace*{0.2cm}
\begin{abstract}
\vspace*{0.2cm}

The analytic forms of the asymptotic quasinormal frequencies of a
coupled scalar field in the Garfinkle-Horowitz-Strominger dilaton
spacetime is investigated by using the monodromy technique
proposed by Motl and Neitzke. It is found that the asymptotic
quasinormal frequencies depend not only on the structure
parameters of the background spacetime, but also on the coupling
between the scalar fields and gravitational field. Moreover, our
results show that only in the minimal couple case, i.e., $\xi$
tends zero, the real parts of the asymptotic quasinormal
frequencies agrees with the Hod's conjecture, $T_H\ln{3}$.
\end{abstract}

 \maketitle

\vspace*{0.2cm}
\section{Introduction}
\vspace*{0.2cm}

It is well known that quasinormal modes possess a discrete spectra
of complex characteristic frequencies which are entirely fixed by
the structure of the background spacetime and are irrelevant of
the initial perturbations\cite{Chandrasekhar}. Thus, one can
directly identify a black hole existence through comparing the
quasinormal modes with the gravitational waves observed in the
universe. Meanwhile, it is generally believed that the study of
the quasinormal modes may lead to a deeper understanding of black
holes and quantum gravity because that the  quasinormal frequency
spectra is related to the AdS/CFT correspondence, string theory
and loop quantum gravity\cite{Hod98}-\cite{Zerilli70}. Therefore,
much attention has been devoted to the study of the quasinormal
modes in the recent thirty years\cite{Detweiler}-\cite{Bachelot}.
In the Schwarzschild spacetime, one found that the asymptotic
quasinormal frequencies of high overtones are described by
\begin{eqnarray}
 \frac{ 2\pi\omega}{\kappa}=\ln{3}+i(2n+1)\pi,\ \ \ n\rightarrow \infty, \label{w1}
 \end{eqnarray}
 where $\kappa $ is the surface gravity constant of the black hole.
 Formula (\ref{w1}) was derived numerically\cite{Nollert} and subsequently
 confirmed analytically\cite{Andersson}\cite{Motl03}.

 Hod\cite{Hod98} first conjectured that the real parts of the
 asymptotic quasinormal frequencies
 of a Schwarzschild black hole can be expressed as $\omega_R=T_H\ln{3}$.
 Together with
 Bohr's correspondence principle, the first law of black hole thermodynamics and the
 asymptotic quasinormal modes, he also obtained some new information
 about the quantization of area at a black hole event horizon. Using of
 Hod's conjecture, Dreyer\cite{Dreyer03} found that the quasinormal
 modes can entirely fixed
 the Barbero-Immirzi parameter\cite{Immirzi57}, which was introduced
 as an indefinite factor by
Immirzi to obtain the right form of the black hole entropy in the
loop quantum gravity. Most significantly, the presence of $\ln{3}$
also means that the gauge group in the loop quantum gravity should
be $SO(3)$ rather than $SU(2)$. Thus, one suggested that Hod's
conjecture maybe create a new way to probe the quantum properties
of black hole.

However, the question whether Hod's conjecture applies to more
general black holes still remain open. Recently, we probed the
asymptotic quasinormal modes of a massless scalar field in the
Garfinkle-Horowitz-Strominger dilaton spacetime\cite{04} and find
that the frequency spectra formula satisfies Hod's conjecture.
Cardoso and Abdalla \cite{Cardoso}\cite{Abdalla} found the
asymptotic quasinormal frequencies in the Schwarzschild de Sitter
and Anti-de Sitter spacetimes depend on the cosmological constant.
Only under the condition that the cosmological constant vanishes,
the real parts of the asymptotic quasinormal frequencies returns
to  $T_H\ln{3}$. For the Reissner-Nordstr\"{o}m black hole, L.
Motl and A.Neitzke\cite{Motl03} obtained the asymptotic
quasinormal frequencies is relevant of the electric charge $Q$. It
is unfortunately that the asymptotic quasinormal frequencies do
not return returns $T_H\ln{3}$ as the black hole charge $Q$ tends
to zero. Thus, some authors\cite{Berti} suggested that Hod's
conjecture should be modified in some way. However, how does the
correct modification look like? It is an interesting subject need
to be study more deeply in the future. At present, it is necessary
and important to study the asymptotic quasinormal modes in the
more general background spacetimes.

In this paper, our main purpose is to investigate the asymptotic
quasinormal modes of a coupled scalar field in the
Garfinkle-Horowitz-Strominger dilaton spacetime. We find that
besides dependence on the structure parameters of the background
spacetime, the asymptotic quasinormal frequencies also are
relevant of the couple constant $\xi$. The plan of the paper is as
follows. In Sec.II, we derive analytically the asymptotic
quasinormal frequency formula of a coupled scalar field in the
Garfinkle-Horowitz-Strominger dilaton spacetime by making use of
the monodromy method\cite{Motl03}. At the last, a summary and some
discussions are presented.

\section{Asymptotic quasinormal frequencies formula of a coupled scalar field
in the Garfinkle-Horowitz-Strominger dilaton spacetime}
\vspace*{0.2cm}

In standard coordinates, the metric for the
Garfinkle-Horowitz-Strominger dilaton black hole spacetime can be
expressed as
 \cite{Garfinkle91}
 \begin{eqnarray}
 ds^2=-\left(1-\frac{2M}{r'}\right)dt^2+\left(1-\frac{2M}{r'}\right)^{-1}
 dr^2+r'(r'-2a)d\Omega^2,\label{gem1}
 \end{eqnarray}
 \begin{eqnarray}
 e^{-2\phi}&=&e^{-2\phi_0}\left(1-\frac{2a}{r'}\right),\nonumber
 \end{eqnarray}
 where $M$ represents the black hole mass and $a$ is a parameter related to dilaton field.
 The dilaton field is given by $e^{-2\phi}=e^{-2\phi_0}(1-\frac{Q^2}{Mr'})$, where $\phi_0$
 is the dilaton value at $r'\rightarrow \infty$ and $Q$ is the electric charge carried by this
 black hole. The relationship among mass $M$, the charge $Q$ and $a$ is described as
 $a=\frac{Q^2}{2M}$.
 This black hole has an event horizon at $r'=2M$ and
 two singular points at $r'=0$ and $r'=2a$.
 The Hawking temperature $T_H=\frac{1}{8\pi M}$ is the same as that of
 the Schwarzschild spacetime.

 In order to simplify the calculation, we introduce a coordinate change
 \begin{eqnarray}
  r=\sqrt{r'(r'-2a)}.
 \end{eqnarray}
Then the metric (\ref{gem1}) can be rewritten as
 \begin{eqnarray}
 ds^2=-\left(1-\frac{2M}{a+\sqrt{a^2+r^2}}\right)dt^2+
 \left(1-\frac{2M}{a+\sqrt{a^2+r^2}}\right)^{-1}\frac{r^2}
 {r^2+a^2}dr^2+r^2d\Omega^2.\label{gem2}
 \end{eqnarray}
The event horizon of the black hole is now located at
$r=2\sqrt{M(M-a)}$ and the Hawking temperature is still described
by $T_H=\frac{1}{8\pi M}$. By means of the quantity
 $R_{\alpha\beta\gamma\delta}R^{\alpha\beta\gamma\delta}$, we
 find  that the point $r=0$ is a curvature singular point.

 The general perturbation equation for a coupled massless scalar field
 in the dilaton spacetime is given by \cite{Frolov}
 \begin{eqnarray}
 \frac{1}{\sqrt{-g}}\partial_\mu(\sqrt{-g}g^{\mu\nu}
 \partial_\nu)\psi-\xi R\psi=0,\label{eq1}
 \end{eqnarray}
where $\psi$ is the scalar field and $R$ is the Ricci scalar
curvature. The coupling between the scalar field and the
gravitational field represented by the term $\xi R\psi$, where
 $\xi$ is a numerical couple factor.

After adopting WKB approximation $\psi=\frac{e^{-i\omega
t}\phi(r)}{r}Y(\theta,\varphi)$, introducing a tortoise coordinate
  \begin{eqnarray}
  x=\sqrt{a^2+r^2}-a+2M \ln{\left[\frac{\sqrt{a^2+r^2}-(2M-a)}
  {2(M-a)}\right]},\label{x2}
  \end{eqnarray}
and substituting Eqs.(\ref{gem2}) and (\ref{x2}) into
Eq.(\ref{eq1}), we know that the radial perturbation equation for
a coupled scalar field in the Garfinkle-Horowitz-Strominger
dilaton spacetime can be expressed as
 \begin{eqnarray}
  \frac{d^2\phi}{dx^2}+(\omega^2-V[r(x)])\phi=0,\label{e3}
 \end{eqnarray}
 where
 \begin{eqnarray}
  V[r(x)]&=&\left(1-\frac{2M}{a+\sqrt{a^2+r^2}}\right)\times\nonumber\\ &&\left[\frac{l(l+1)}{r^2}+
  \frac{2M(a^2+r^2)^{3/2}-2\sqrt{a^2+r^2}a^3+2Ma^3-2a^4-r^2a^2}
  {r^4(a+\sqrt{r^2+a^2})^2}+\xi R\right]\label{v},
 \end{eqnarray}
 and
 \begin{eqnarray}
 R=\frac{2a^2(r^2+2aM-2M\sqrt{a^2+r^2})}{r^6}.
 \end{eqnarray}
It is well known that the quasinormal modes consist of the
solutions of the perturbation equation (\ref{e3}) with the
boundary conditions appropriate for purely ingoing waves at the
event horizon and purely outgoing waves at infinity, namely,
 \begin{eqnarray}
  \phi &=& e^{+i\omega x},\ \ \ \ \ \ x\rightarrow -\infty, \nonumber\\
  \phi &=& e^{-i\omega x},\ \ \ \ \ \ x\rightarrow +\infty.
  \end{eqnarray}
In general, we just consider the perturbation equation (\ref{e3})
in the physical region $r\geq 2\sqrt{M(M-a)}$ in the
Garfinkle-Horowitz-Strominger dilaton black hole. However, in the
monodromy method, it is fundamental to extend analytically
Eq.(\ref{e3}) to the whole complex $r$-plane. In the process of
analytical extension, we find that both the tortoise coordinate
$x(r)$ and the wave function $\phi(r)$ are multivalued around the
singular points $r=0$ and $r=2\sqrt{M(M-a)}$ . This
multivaluedness plays an important and essential role in our
analysis. As in Ref.\cite{Motl03}, we can put branch cuts in the
complex $r$-plane from $r=0$ to $r=2\sqrt{M(M-a)}$ in order to
avoid dealing with multivalued functions. The monodromy of
$\phi(r)$ can be defined by the discontinuity across the cut.
Finally, by comparing the local and global monodromy of $\phi(r)$
along the selected contour $L$ around the point
$r=2\sqrt{M(M-a)}$, we can obtain the asymptotic quasinormal
frequency spectra in the Garfinkle-Horowitz-Strominger dilaton
black hole spacetime.

From Eq.(\ref{x2}), we find that $x$ is not uniquely defined as a
function of $r$. However, it is very fortunate that we can
determine the sign of $Re(x)$ in the complex $r$-plane. The
regions for the different sign of $Re(x)$ are shown in the Figure
1.
\begin{figure}[ht]
\begin{center}
\includegraphics[width=6cm]{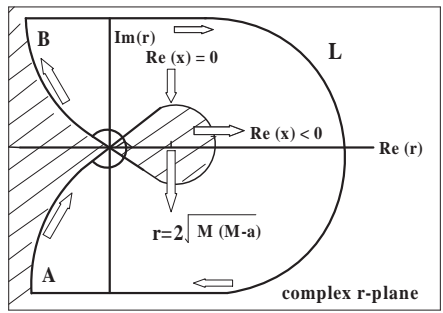}
\caption{The complex $r$-plane and the contour $L$. The regions
with the hachures denote the area $Re(x)<0$.}
\end{center}
\label{fig1}
\end{figure}
As in Ref.\cite{Motl03}, in order to compute conveniently, we may introduce the
variable $z=x-\frac{\pi i}{2\kappa}$.  For $r=0$, we have $z=0$. To fix
the angle of the variable $z$ at the point $r=0$, we
define the branch $n=0$ for $\ln{(-1)}$.

Now, we must define the boundary condition at $r=\infty$.
Similarly to Ref.\cite{Motl03}, we can analytically continue
$\phi(r)$ via ``Wick rotation" to the line $Im(\omega x)=0$. For
the highly damped modes, i.e., $\omega$ are almost purely
imaginary, the line $Im(\omega x)=0$ is just slightly sloped off
the line $Re(x)=0$. Assuming initially that $Re(\omega)>0$ and
then making use of the condition $Im(\omega x)=0$, we can obtain
$x=+\infty$ is rotated to $\omega x=+\infty$. Thus, on the line
$Re(x)=0$, the boundary condition at $r=+\infty$ actually becomes
 \begin{eqnarray}
 \phi(r)\sim e^{-i\omega x},\ \ \ \ \omega x\rightarrow +\infty. \label{b1}
 \end{eqnarray}

Let us now compute the local monodromy around the singular point
 $r=2\sqrt{M(M-a)}$. This can be done by
 matching the asymptotic along the line $Re(x)=0$, i.e., the contour $L$
 shown in the Fig. 1. When we
 start at point $A$ and move along the contour $L$ towards interior, the $\phi(x)$
 can be look as the plane waves because the term $\omega^2$ dominates the
 potential in Eq.(\ref{e3}) away from the origin point.
 At the vicinity of the point $r=0$, we have
 \begin{eqnarray}
 z\sim -\frac{r^2}{2(M-a)},
 \end{eqnarray}
 and the behaviors of the Ricci scalar curvature and the potential are
 \begin{eqnarray}
 R\sim \frac{2a(a-M)}{r^4},
 \end{eqnarray}
 and
 \begin{eqnarray}
 V[r(z)]\sim-\frac{1-2\xi}{4z^2}.\label{v4}
 \end{eqnarray}
 We make the identification $j=\sqrt{2 \xi}$, and then the perturbation
 equation (\ref{e3}) can be rewritten as
 \begin{eqnarray}
 (\frac{d^2}{dz^2}+\omega^2+\frac{1-j^2}{4z^2})\phi(z)=0.
 \end{eqnarray}
 From the Ref.\cite{handbook}, we find it can be exactly solved in terms of the
 Bessel function and the general solution near the origin point can be expressed as
 \begin{eqnarray}
 \phi(z)=A_+c_+\sqrt{\omega z}J_{+j/2}(\omega z)+
 A_-c_-\sqrt{\omega z}J_{-j/2}(\omega z).\label{J1}
 \end{eqnarray}
 Now, let us look for the asymptotic forms of the solution (\ref{J1}) away from the origin.
 After considering the asymptotic behavior of $J_{\pm j/2}(\omega z)$ as
 $\omega z\rightarrow \infty$, we can select the normalization factors $c_{\pm}$
 in (\ref{J1}) so that we can write the asymptotic forms as
\begin{eqnarray}
 c_{\pm}\sqrt{\omega z}J_{\pm \frac{j}{2}}(\omega z)\sim 2\cos{(\omega z-\alpha_{\pm})},
 \label{J2}
 \end{eqnarray}
  as $\omega z\rightarrow \infty$ , where $\alpha_{\pm}=\frac{\pi}{4}(1\pm j)$.
  From Eqs.(\ref{J1}), (\ref{J2}) and the boundary condition (\ref{b1}), we have
  \begin{eqnarray}
  A_+e^{-i\alpha_+}+A_-e^{-i\alpha_-}=0,
  \end{eqnarray}
 and
  \begin{eqnarray}
  \phi(z)\sim (A_+e^{i\alpha_+}+A_-e^{i\alpha_-})e^{-i\omega z}.
   \end{eqnarray}
To follow the contour $L$ and approach to the point $B$, we have
to turn an angle $\frac{3\pi}{2}$ around the origin $r=0$,
corresponding to $3\pi$ around $z=0$. From the Bessel function
   behavior near the origin point
    \begin{eqnarray}
   J_{\pm \frac{j}{2}}(\omega z)=z^{\pm \frac{j}{2}}\varphi(z),
   \end{eqnarray}
   where $\varphi(z)$ is an even holomorphic function, we find that after the $3\pi$
   rotation the asymptotic are
   \begin{eqnarray}
 c_{\pm}\sqrt{\omega z}J_{\pm \frac{j}{2}}(\omega z)\sim e^{6i\alpha_{\pm}}
 2\cos{(-\omega z-\alpha_{\pm})},
 \label{J3}
 \end{eqnarray}
  as $\omega z\rightarrow -\infty$. Thus the asymptotic at the point $B$
  \begin{eqnarray}
 \phi(z)\sim (A_+e^{5i\alpha_+}+A_-e^{5i\alpha_-})e^{-i\omega z}+
 (A_+e^{7i\alpha_+} +A_-e^{7i\alpha_{-}})e^{i\omega z},\ \ \ \  \omega z\rightarrow -\infty.
 \label{J4}
 \end{eqnarray}
 Finally, we can come back from the point $B$ to the point $A$ along the large semicircle
 in the right half-plane.
 In this region, because the term $\omega^2 $ dominates the potential $V[r(x)]$, we can
 approximate the solutions of the perturbation equation as plane waves. When we return to the
  point $A$,
 the coefficient of $e^{-i\omega z}$ remains unchange. While the coefficient of
  $e^{i\omega z}$ makes only an exponentially small contribution to $\phi(z)$
  in the right plane. Finally, we find that the monodromy around the contour $L$ must
  multiply the coefficient of $e^{-i\omega z}$ by a factor
  \begin{eqnarray}
  \frac{A_+e^{5i\alpha_+}+A_-e^{5i\alpha_-}}{A_+e^{i\alpha_+} +A_-e^{i\alpha_{-}}}
  =\frac{e^{6i\alpha_+}-e^{6i\alpha_-}}{e^{2i\alpha_+} -e^{2i\alpha_{-}}}=-(1+2\cos{\pi
  j}).
 \label{11}
  \end{eqnarray}

 Now, let us calculate the global monodromy around the contour $L$. Since the only
 singularity of $\phi(r)$ or $e^{-i\omega z}$ inside the contour occurs at the
  point $r=2\sqrt{M(M-a)}$, according to the boundary condition of the
  quasinormal modes, we can obtain the monodromy of $\phi(r)$ or
   $e^{-i\omega z}$ at this point. After a full clockwise round trip, $\phi(r)$ acquires a
   phase $e^{\frac{\pi\omega}{\kappa}}$, while $e^{-i\omega z}$ acquires a phase
    $e^{-\frac{\pi\omega}{\kappa}}$. So the coefficient of
  $e^{-i\omega z}$ in the asymptotic of $\phi(r)$ must be multiplied by
  $e^{\frac{2\pi\omega}{\kappa}}$. Substituting $j=\sqrt{2\xi}$ into Eq.(\ref{11}) and
  comparing the local monodromy with the global one, we find the
  asymptotic quasinormal frequencies of a coupled scalar field in the Garfinkle-Horowitz-Strominger
  dilaton black hole spacetime satisfy
  \begin{eqnarray}
     e^{\frac{2\pi\omega}{\kappa}}=-[1+2\cos{(\sqrt{2\xi} \pi)}]\label{22}.
     \end{eqnarray}
Making a simple operation on Eq.(\ref{22}), we obtain easily the
formula
  \begin{eqnarray}
       \frac{2\pi\omega}{\kappa}=\ln{[1+2\cos{(\sqrt{2\xi} \pi)}]}+i(2n+1)\pi,\ \ \ \
       n \rightarrow \infty.\label{12}
     \end{eqnarray}
It is interesting to note that the right-hand side of the formula
(\ref{12}) contains the couple factor $\xi$. This means that the
asymptotic quasinormal modes depend  not only on the structure
parameters of the background spacetimes, but also on the coupling
between the  matter fields and gravitational field. Furthermore,
we find only when $\xi=0$, namely, in the minimal  couple case,
the real part of the right-hand side of the formula (\ref{12})
becomes $\ln{3}$, which is consistent with Hod's conjecture.

  \section{Summary and discussion}
  \vspace*{0.2cm}

We have investigated the analytical forms of the asymptotic
quasinormal frequencies for a coupled scalar field in the
Garfinkle-Horowitz-Strominger dilaton spacetime by adopting the
monodromy technique. It is shown that the asymptotic quasinormal
frequencies depend not only on the structure parameters of the
background spacetime, but also on a couple constant $\xi$. The
fact tells us that the interaction between the matter fields and
gravitational field will affect the frequencies spectra formula of
the asymptotic quasinormal modes. It is a novel property of the
quasinormal modes in Garfinkle-Horowitz-Strominger dilaton
spacetime. In the Schwarzschild and Reissner-Nordstr\"{o}m
spacetimes, we find the asymptotic quasinormal frequencies do not
possess this behavior because the curvature scalar $R$ in both the
spacetimes is equal to zero and then the coupled term in
Eq.(\ref{eq1}) vanishes. Moreover, we find that Hod's conjecture,
the real parts of the asymptotic quasinormal frequencies equals to
$T _H \ln{3}$, is valid only for the minimal couple case in the
Garfinkle-Horowitz-Strominger dilaton spacetime, i.e., $\xi$
becomes zero. It implies that there maybe exist a more general
form of the Hod's conjecture.

It should be pointed out that although the formula (\ref{12}) is
not related to the dilaton field parameter $a$ obviously, it does
not means that the quasinormal modes are independent of the
dilaton. The reason is that we just consider the contribution of
the leading term $r^{-4}$ in the potential $V$ to the local
monodromy of the wave function $\phi(z)$. If we consider the
contribution of the lower order  terms in the potential $V(r)$,
the corrected term to the asymptotic quasinormal frequencies will
depend on the parameter $a$ of the dilaton field. For example,  as
$a$ is very small, the main corrected term is roughly in
proportion to $\frac{(1-i)}{\sqrt{n+1/2}}[
l(l+1)+1-\frac{2M}{3(M+a)}]$, which shows that the quasinormal
frequencies increase as $a$ increases in the
Garfinkle-Horowitz-Strominger dilaton spacetime.

\begin{acknowledgments}
This work was supported by the National Natural Science Foundation
of China under Grant No. 10275024; the FANEDD under Grant No.
200317; and the Hunan Provincial Natural Science Foundation of
China under Grant No. 04JJ3019.
\end{acknowledgments}

\end{document}